\documentclass[aps,prl,superscriptaddress,amsmath,amssymb,twocolumn,amsfonts,floatfix, longbibliography]{revtex4-2}

\usepackage{graphicx}
\usepackage{epstopdf}
\usepackage[T1]{fontenc}
\usepackage[latin9]{inputenc}
\usepackage{amsbsy}
\usepackage{gensymb}
\usepackage{amsmath}
\usepackage{amssymb}
\usepackage{bbm}
\usepackage{braket}
\allowdisplaybreaks

\usepackage[dvipsnames]{xcolor}
\definecolor{deepfuchsia}{rgb}{0.76, 0.33, 0.76}
\definecolor{electricpurple}{rgb}{0.75, 0.0, 1.0}
\usepackage[colorlinks=true,linktoc=page,linkcolor=blue,citecolor=magenta,urlcolor=electricpurple]{hyperref}

\usepackage[normalem]{ulem}
\usepackage{orcidlink}

\newcommand{\figref}[1]{Fig.~\ref{#1}}

\newcommand{\tableref}[1]{Table~\ref{#1}}

\begin{document}

\title{\textrm{Nonsymmorphic symmetry protected hourglass Dirac chain topology and conventional superconductivity in ZrIrGe}}


\author{{Pavan Kumar Meena}\,\orcidlink{0000-0002-4513-3072}}\thanks{These authors contributed equally to this work}
\affiliation{Department of Physics, Indian Institute of Science Education and Research Bhopal, Bhopal, 462066, India}
\author{{Dibyendu Samanta}\,\orcidlink{0009-0004-3022-7633}}\thanks{These authors contributed equally to this work}
\affiliation{Department of Physics, Indian Institute of Technology, Kanpur 208016, India}
\author{{Shashank Srivastava}}
\affiliation{Department of Physics, Indian Institute of Science Education and Research Bhopal, Bhopal, 462066, India}
\author{{Poulami Manna}}
\affiliation{Department of Physics, Indian Institute of Science Education and Research Bhopal, Bhopal, 462066, India}
\author{{Sudeep Kumar Ghosh}\,\orcidlink{0000-0002-3646-0629}}
\email[]{skghosh@iitk.ac.in}
\affiliation{Department of Physics, Indian Institute of Technology, Kanpur 208016, India}
\author{{Ravi Prakash Singh}\,\orcidlink{0000-0003-2548-231X}}
\email[]{rpsingh@iiserb.ac.in}
\affiliation{Department of Physics, Indian Institute of Science Education and Research Bhopal, Bhopal, 462066, India}

\begin{abstract}
Ternary transition-metal germanide superconductors with nonsymmorphic symmetries offer promising platforms for symmetry-protected topological phases. In this work, we investigate ZrIrGe, which crystallizes in the nonsymmorphic TiNiSi-type structure. Electrical, magnetic, and specific heat measurements confirm bulk type-II superconductivity with a full gap and a transition temperature of 2.84(7) K, consistent with weak-coupling BCS behavior. First-principles calculations reveal hourglass-shaped bulk band dispersions and a Dirac chain composed of symmetry-protected fourfold-degenerate Dirac points, leading to drumhead-like surface states near the Fermi level. Additionally, ZrIrGe exhibits a nontrivial $\mathbb{Z}_2$ topological character, resulting in helical surface states that cross the Fermi level, making it a strong candidate for proximity-induced topological superconductivity. The coexistence of conventional superconductivity and topological band features establishes ZrIrGe as a rare stoichiometric system for exploring intrinsic topological superconductivity.
\end{abstract}
\maketitle

\section{INTRODUCTION}
The search for intrinsic topological superconductivity has gained considerable momentum due to its potential to host Majorana modes and its relevance for topological quantum computing~\cite{qi2011topological, sato2017topological}. While proximity-induced superconductivity in doped topological insulators such as Bi$_2$Se$_3$ and SnTe~\cite{wang2012coexistence, sasaki2012odd} has achieved some success, these approaches often face challenges related to disorder and interface effects. This has motivated the search for stoichiometric materials that naturally combine superconductivity and nontrivial topology-without external tuning~\cite{yan2013large, guan2016superconducting}. Among the most promising candidates are topological metals with nonsymmorphic crystal symmetries, where space group operations such as glide mirrors and screw axes enforce symmetry-protected band crossings~\cite{Young2015, Zhao2016}. These can give rise to exotic fermionic excitations, such as hourglass fermions~\cite{wang2016hourglass,Singh2018,gao2020r,Li2018,ma2017experimental}. However, intrinsic topological superconductivity in these systems remains rare, often limited by low superconducting transition temperatures ($T_c$) and competing bulk metallic states. Nevertheless, bulk superconductors that intertwine symmetry-protected topology, strong spin-orbit coupling (SOC), and unconventional pairing mechanisms continue to offer fertile ground for discovering new quantum phases.  

Among such candidates, ternary transition-metal silicides and germanides ~\cite{gupta2015review, morozkin1999crystallographic, landrum1998tinisi, subba1985structure}, with the orthorhombic TiNiSi-type structure, have recently attracted attention for exhibiting both superconductivity and nontrivial band topology. Several members of this family show fully gapped, nodeless superconductivity~\cite{tay2023nodeless, kp2018superconducting, kp2020probing}, and intriguingly, time-reversal symmetry (TRS) breaking has been reported in some compounds but not in others~\cite{panda2019probing, bhattacharyya2021electron, ghosh2022time}. These contrasting behaviors raise fundamental questions about their pairing mechanisms. Theoretical studies have suggested that compounds within this class could host intrinsic topological superconductivity~\cite{panda2024probing, uzunok2020first}, with nonsymmorphic symmetries playing a critical role in stabilizing symmetry-protected hourglass dispersions, as recently predicted in (Ti, Hf)IrGe \cite{meena2025topological}. These materials are structurally related to U(Rh, Co)Ge, which exhibits ferromagnetic superconductivity, possibly of triplet nature~\cite{hattori2012superconductivity}. Furthermore, the parent binary compound IrGe is known to host soft phonon modes that may facilitate unconventional pairing~\cite{arushi2022microscopic}. However, its equiatomic analogue ZrIrGe remains largely unexplored despite its close chemical and structural similarity. With a SOC strength characteristic of 4$d$ elements, ZrIrGe is expected to exhibit intermediate band splittings between those of Ti (3$d$) and Hf (5$d$), offering a promising platform to systematically tune and correlate superconducting properties with SOC via atomic substitution.

\begin{figure*}[!ht]
\includegraphics[width=2.05\columnwidth]{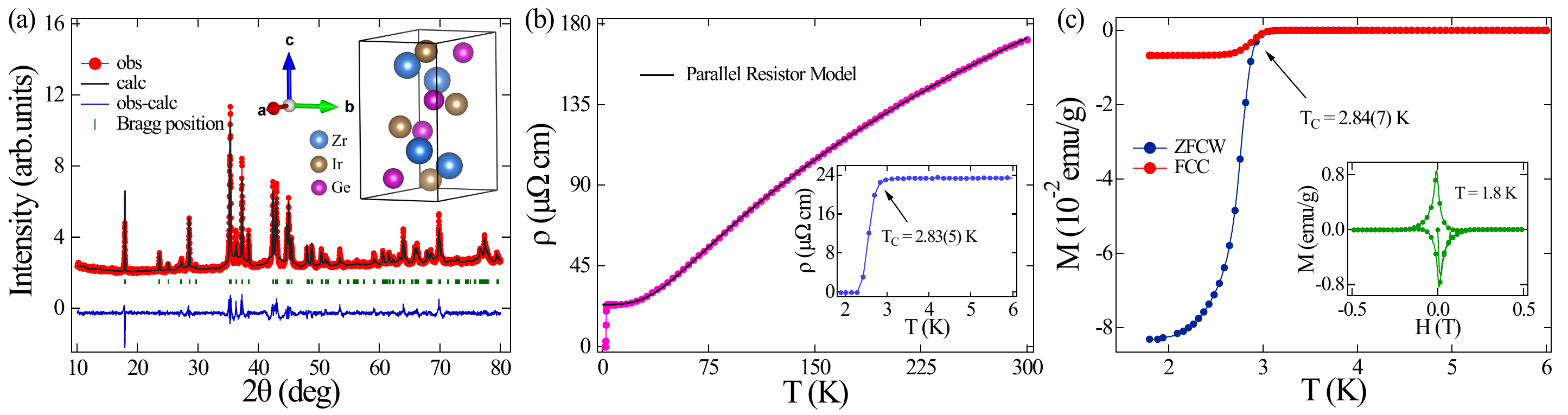}
\caption {\label{Fig1} \textbf{Structural and bulk superconductivity characterization of ZrIrGe:} (a) Powder XRD patterns of ZrIrGe with Rietveld refinement, where the red markers represent experimental data, the black line indicates the theoretical refinement, and the vertical bars correspond to the Bragg positions. The line at the bottom depicts the difference between the observed and calculated data. The inset represents the TiNiSi-type crystal structure of the ZrIrGe compound. (b) Temperature-dependent electrical resistivity [$\rho(T)$] measured at H = 0 mT for ZrIrGe, with inset showing an enlarged view of $\rho(T)$, confirming the superconductivity at $T_{\rm c}$ = 2.83(5) K. (c) Magnetization as a function of temperature at an applied magnetic field H = 1.0 mT, measured in ZFCW and FCC mode, confirming superconductivity. The inset depicts the magnetic field-dependent magnetization at 1.8 K.}
\end{figure*}

In this study, we systematically investigate the superconducting and topological properties of ZrIrGe, an Ir-based equiatomic ternary germanide \cite{wang1987crystal}, as a promising candidate for intrinsic topological superconductivity. Electrical resistivity, magnetization, and specific heat measurements confirm bulk type-II superconductivity with a transition temperature of $T_c \sim 2.84$ K, consistent with fully gapped, weak-coupling BCS behavior. First-principles calculations and symmetry analysis reveal hourglass-type bulk dispersions, where the necks form Dirac rings protected by nonsymmorphic symmetries, leading to drumhead-like surface states. Additionally, ZrIrGe hosts well-separated Dirac topological surface states with helical spin textures, arising from a nontrivial $\mathbb{Z}_2$ topology. These robust, symmetry-protected states disperse across the Fermi level while remaining distinct from the bulk states-features that are further supported by the material's structural similarity to (Ti, Hf)IrGe, known to exhibit hourglass fermions \cite{meena2025topological}. Together, these findings position ZrIrGe as a rare stoichiometric system in which conventional superconductivity coexists with symmetry-enforced topological band features, offering a valuable platform for probing the interplay between superconducting order and topological surface states.

\section{Methods}

Polycrystalline ZrIrGe samples were synthesized by arc melting a stoichiometric amount of high-purity ($4N$) Zr, Ir, and Ge under an argon atmosphere. Samples were remelted multiple times for homogeneity. Phase purity and crystal structure were determined by powder X-ray diffraction (XRD) using a PANalytical X'pert Pro diffractometer equipped with Cu $K_{\alpha}$ radiation with $\lambda$ = 1.5406 $\text{\AA}$. Magnetization measurements were performed using a Quantum Design MPMS-7T. Electrical responses were measured based on the four-probe approach using a Quantum Design PPMS-9T. Specific heat measurements were performed with a Quantum Design PPMS instrument equipped with a dilution refrigerator.

We performed \textit{ab initio} electronic structure calculations of ZrIrGe using density functional theory (DFT) within the QUANTUM ESPRESSO package~\cite{Giannozzi2009}. Exchange-correlation effects were handled using the generalized gradient approximation (GGA) with the Perdew-Burke-Ernzerhof (PBE) functional~\cite{Perdew1996}. To model the electron-ion interactions, we employed projector-augmented wave (PAW) pseudopotentials. The plane-wave basis set was truncated at a kinetic energy cutoff of 80 Ry. For Brillouin zone sampling, we used a $8 \times 10 \times 8$ Monkhorst-Pack k-point grid for bulk calculations. All calculations were based on experimental lattice constants and atomic positions obtained from Rietveld refinement of X-ray diffraction data, as detailed below.

\begin{figure*}
\includegraphics[width=2.05\columnwidth]{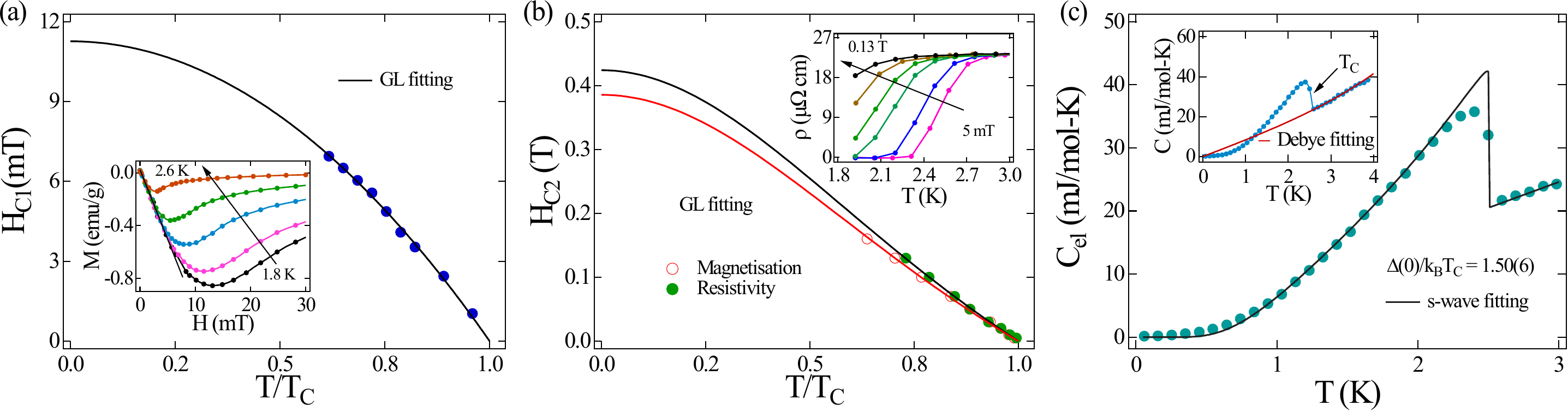}
\caption {\label{Fig2} \textbf{Lower and upper critical fields and specific heat analysis of ZrIrGe:} (a) Temperature dependence of the lower critical field $H_{c1}$, fitted using the Ginzburg-Landau (GL) relation, with the inset illustrating the low-field magnetization behavior at various isotherms. (b) The upper critical field $H_{c2}$, was determined from resistivity and magnetization measurements, fitted to the GL relation. The inset presents resistivity curves at different magnetic fields for ZrIrGe. (c) The temperature-dependent electronic specific heat C$_{el}$, obtained after subtracting the phonon contribution, fitted with isotropic s-wave model. The inset showing specific heat C/T vs $T^{2}$ plot confirms superconductivity, with the data in the normal state well-fitted using the Debye model, allowing extraction of the phonon contribution.}
\end{figure*}

\section{RESULTS AND DISCUSSION}

\subsection{Sample characterization}

The refined powder X-ray diffraction (XRD) pattern of the ternary compound ZrIrGe, depicted in \figref{Fig1}(a), confirms that it crystallizes in the orthorhombic TiNiSi-type structure with the space group $Pnma$ (no. 62) and the point group $D_{2h}$. The corresponding crystal structure is shown in the inset of \figref{Fig1}(a). Structural determination was performed through Rietveld refinement using FullProf Suite software \cite{rodriguez1993recent}. The lattice constants of ZrIrGe are $a=6.5687(7) \text{\AA}$, $b=4.0275(1) \text{\AA}$, and $c=7.5280(1) \text{\AA}$ in good agreement with previous studies \cite{wang1987crystal}. Zr, Ir, and Ge atoms occupy the same 4$c$ Wyckoff position with fractional coordinates $(0.027, 0.25, 0.675)$, $(0.148, 0.25, 0.064)$, and $(0.258, 0.25, 0.376)$, respectively.


\subsection{Bulk superconducting properties}

Electrical resistivity $\rho(T)$ as a function of temperature was measured for the ternary compound ZrIrGe in a zero magnetic field ($H = 0$ mT). As shown in \figref{Fig1}(b), $\rho(T)$ exhibits metallic behaviour, decreasing as the temperature decreases. The residual resistivity ratio (RRR), defined as $\rho_{300 K}/\rho_{10 K}$, was determined to be 7.31(5), which is higher than that of structurally similar compounds \cite{kase2016superconductivity}, indicating the good quality of the polycrystalline sample. At the critical temperature $T_c$ of 2.83(5) K, $\rho(T)$ drops sharply and approaches zero, consistent with previous reports \cite{wang1987crystal}. The inset of \figref{Fig1}(b) provides a zoomed-in view of the $\rho(T)$ behavior near $T_c$. The normal-state resistivity data were nicely fitted using the theoretical Wiesmann parallel resistivity model \cite{wiesmann1977simple}, expressed as:
\begin{equation}
\frac{1}{\rho(T)}= \frac{1}{\rho_{s}}+ \frac{1}{\rho_{i}(T)}.
\label{Eq1:Parallel}
\end{equation}
where $\rho_{s}$ represents the temperature-independent saturated resistivity, corresponding to a mean free path of the order of interatomic spacing \cite{fisk1976saturation}. The ideal contribution to resistivity, $\rho_{i}(T)$, can be written as the sum of two terms according to Matthiessen's rule: $\rho_{i}$ = $\rho_{i,0}$+$\rho_{i,L}$. Here, $\rho_{i,o}$ represents the residual resistivity measured due to impurity scattering, and $\rho_{i, L}$ represents the temperature-dependent contribution caused by thermally excited phonons given by:
\begin{equation}
{\rho_{i,L}(T)}=C \left(\frac{T}{\Theta_{D}}\right)^p \int_{0}^{{\Theta_{D}}/T} \frac{x^{p}}{(e^{x}-1)(1-e^{-x})} dx.
\label{Eq2:Parallel}
\end{equation}
Here, $\Theta_{D}$ is the Debye temperature, C is a material-specific constant, and $p$ is an exponent that depends on the dominant interaction mechanism. For example, $p = 3$ corresponds to Wilson's theory, while $p = 5$ aligns with the BG formula \cite{grimvall1981electron}. Fitting the resistivity data using $p = 3$ yielded $\rho_{0}$ = 24.61 $\mu\ohm$cm, $\rho_{s}$ = 514.44 $\mu\ohm$cm, and $\Theta_{D}$ = 272.1 K. These results align with the trends observed in related ternary silicides (MIrSi) \cite{kase2016superconductivity}.


Temperature-dependent magnetization data were collected in two modes: zero field cooled warming (ZFCW) and field cooled cooling (FCC), under an applied field of 1.0 mT, as presented in Fig.~\ref{Fig1}(c) for ZrIrGe. Magnetization data revealed a diamagnetic nature below the superconducting transition temperature ($T_{\rm c}$) of 2.84(7) for ZrIrGe, which is consistent with resistivity measurement. Difference FCC and ZFCW mode data suggest strong flux pinning confirming type-II superconductivity in the ZrIrGe compound. High-field magnetization data at 1.8 K, shown in the inset of Fig.~\ref{Fig1}(c), further confirmed the type-II nature of superconductivity. The magnetization data are presented without correction for the sample's demagnetizing field \cite{aharoni1998demagnetizing}.
  
To determine the lower critical field $H_{c1}(0)$, magnetization as a function of the applied field, M(H), was measured at various temperatures. The deviation of the M(H) curve from the linear Meissner line at low fields defines the value of $H_{c1}$ for each isotherm, as illustrated in the inset of Fig.~\ref{Fig2}(a). Temperature-dependent $H_{c1}$ data were analyzed using the Ginzburg-Landau (GL) equation:
\begin{equation}
H_{c1}(T)=H_{c1}(0)\left[1-\left(\frac{T}{T_{\rm c}}\right)^{2}\right].
\label{eqn1:HC1}
\end{equation}
The obtained value of $H_{c1}(0)$ is 11.2(1) mT, and the fitting of the data with this approach is shown in Fig.~\ref{Fig2}(a). Further, upper critical field $H_{c2}$ was extracted from temperature-dependent magnetisation and resistivity measurements under various applied magnetic fields. The resistivity data showed a reduction in the transition temperature $T_{\rm c}$ with increasing magnetic field (H), as depicted in the inset of Fig.~\ref{Fig2}(b). The $H_{c2}(T)$ data were fitted using the GL equation:
\begin{equation}
H_{C2}(T) = H_{C2}(0)\left[\frac{(1-t^{2})}{(1+t^{2})}\right];  \quad  \text{where} \;  t = \frac{T}{T_{\rm c}}.
\label{eqn3:HC1}
\end{equation}
The fitting results yielded $H_{c2}$(0) values of 0.38(1) T from magentisation and 0.42(1) T from resistivity, as shown in Fig.~\ref{Fig2}(b).

Superconductivity can be suppressed by either orbital-limiting or Pauli spin paramagnetic-limiting effects. In the orbital limit, superconductivity is destroyed when the kinetic energy of Cooper pairs exceeds the condensation energy under a magnetic field. For the Pauli-limiting effect, superconductivity breaks down when one of the electrons in a Cooper pair aligns its spin with the applied field. In the absence of Pauli spin paramagnetism and spin-orbit interaction, the Werthamer-Helfand-Hohenberg (WHH) theory provides the orbital limiting upper critical field $H^{orb}_{c2}(0)$ for type-II superconductors, which is given by \cite{werthamer1966temperature, helfand1966temperature}:
\begin{equation}
H^{orb}_{c2}(0) = -\alpha T_{\rm c} \left.{\frac{dH_{c2}(T)}{dT}}\right|_{T=T_{\rm c}}, 
\label{eqn3:WHH}
\end{equation}
where $\alpha$ is the purity factor 0.693 (0.73) for dirty (clean) limit superconductors. With a slope of $\frac{dH_{c2}(T)}{dT}|_{T=T_{\rm c}}$ and $\alpha$ = 0.693 for a dirty limit superconductor, the orbital limiting field was calculated to be $H_{c2}(0)$ = 1.12(6) T for ZrIrGe. The Pauli limiting field $H^P_{c2}$(0) for a BCS superconductor is given by $H^P_{c2}$(0) = 1.86 $T_{\rm c}$, yielding a value of 5.29(5) T for $T_{\rm c}$ = 2.84(7) K \cite{chandrasekhar1962note, clogston1962upper}. The Maki parameter $\alpha_{m}$, which quantifies the relative significance of the orbital and Pauli effects, is defined as $\alpha_{m}= \sqrt{2} H_{c2}^{orb}/H_{c2}^{p}$ \cite{maki1966effect}. The calculated $\alpha_{m}$ of 0.30 (1), less than unity, suggests a minimal influence of the Pauli limiting field.

The GL coherence length $\xi_{GL}$ was estimated using $H_{c2}(0) = \frac{\Phi_{0}}{2\pi\xi_{GL}^{2}}$, where $\Phi_{0}$ = $2.07 \times 10^{15}$ Wb is the flux quantum \cite{tinkham2004introduction}. For $H_{c2}$(0) = 0.42 T, $\xi_{GL}$ was calculated to be 29.4(4) $nm$. The superconducting penetration depth $\lambda_{GL}$(0) was derived from $H_{c1}$(0) and $\xi_{GL}$ using: 
\begin{equation}
H_{c1}(0) = \frac{\Phi_{0}}{4\pi\lambda_{GL}^2(0)}\left[ln \frac{\lambda_{GL}(0)}{\xi_{GL}(0)} + 0.12\right].
\end{equation}
The value of $\lambda_{GL}$(0) was found to be 163.9(3) nm, yielding a GL parameter $\kappa_{GL}$ = $\lambda_{GL}(0)$/$\xi_{GL}(0)$ = 5.5(5). As $\kappa_{GL}$ $>$ 1/$\sqrt{2}$, this confirms that ZrIrGe is a type-II superconductor. Finally, the thermodynamic critical magnetic field $H_{c}$ was estimated using $H_{c}^{2}.ln k_{GL}= H_{c1} H_{c2}$. The derived parameters $\lambda_{GL}$, $\xi_{GL}$, $H_{c}$, and $K_{GL}$ are consistent with those of similar compounds, as shown in \tableref{tbl: parameters}.


Evidence for bulk superconductivity in ZrIrGe compounds was found through zero-field-specific heat measurements, C(T). A clear jump in C(T) (inset of \figref{Fig2}(c)) confirms the superconducting transition at $T_{\rm c} \sim 2.6$ K consistent with previous studies. The normal-state data follows the Debye model: 
\begin{equation}
    C = \gamma_{n} T + \beta_{3} T^{3} + \beta_{5} T^{5},
\end{equation}
where the electronic contribution is given by $\gamma_{n} T$, and the lattice contribution is represented by $\beta_{3} T^{3}$ + $\beta_{5} T^{5}$. The parameters obtained are $\gamma_{n}$ = 8.25(8) mJ/mole K$^{2}$ and $\beta_{3}$ = 0.11(1) mJ/mole K$^{4}$ and $\beta_{5}$ = 0.0013(8) mJ/mole K$^{6}$. Within the Debye approximation, $\beta_{3}$ is related to the Debye temperature $\theta_{D}$ by: $\theta_{D} = \left(\frac{12\pi^{4} R N}{5 \beta_{3}}\right)^{\frac{1}{3}}$, where R = 8.314 J mol$^{-1}$ K$^{-1}$ is the universal gas constant and N = 3 (atoms per formula unit in ZrIrGe). Based on these values, the calculated $\theta_{D}$ for ZrIrGe is 374(9). The electronic density of state at the Fermi level, $D_{C}(E_{F})$, is derived from: $\gamma_{n}$ = $\left(\frac{\pi^{2} k_{B}^{2}}{3}\right)$ $D_{C}(E_{F})$, where $k_{B}$ = 1.38 $\times$ 10$^{-23}$ J K$^{-1}$. The obtained value $D_{C}(E_{F})$ is 3.50 (1) states $eV^{-1}f.u.^{-1}$ for the ZrIrGe. 

The electron-phonon coupling constant, $\lambda_{el-ph}$ is determined using McMillan's equation, which is applicable to conventional phonon-mediated superconductors within the weak-to-intermediate coupling regime \cite{mcmillan1968transition}:
\begin{equation}
\lambda_{e-ph} = \frac{1.04+\mu^{*}\mathrm{ln}(\theta_{D}/1.45T_{\rm c})}{(1-0.62\mu^{*})\mathrm{ln}(\theta_{D}/1.45T_{\rm c})-1.04 }.
\label{eqn8:Lambda}
\end{equation}
Here, $\mu^{*}$ represents the repulsive screened Coulomb interaction, typically ranging from 0.07 to 0.15. $\mu^{*}$ is taken as 0.13 for intermetallic compounds. Using the estimated $\theta_{D}$ and observed $T_{\rm c}$, the calculated $\lambda_{e-ph}$ = 0.51(3) indicates weak electron-phonon coupling in the ZrIrGe compound.

To investigate the symmetry of the superconducting gap, the electronic specific heat, C$_{el}(T)$, is obtained by subtracting the contribution of the lattice, $\beta_{3} T^3 + \beta_{5} T^{5}$, from the total specific heat, C(T). The low temperature C$_{el}(T)$ data are well described by the s-wave model, as illustrated in \figref{Fig2}(c). In the BCS s-wave framework, the entropy S for a single-gap superconductor is given by \cite{padamsee1973quasiparticle}:
\begin{equation}
\!\!\!\!\frac{S}{\gamma_{n} T_{\rm c}}= -\frac{6\Delta(0)}{\pi^{2} k_{B} T_{\rm c}} \int_{0}^{\infty}\!\!\! dy \left[ {fln(f)+(1-f)ln(1-f)} \right], 
\label{BCS}
\end{equation}

where, $f(\xi)$ = ${e^{E(\xi)/k_{B}T}+1}^{-1}$ is the Fermi function and $y$ = $\xi/\Delta(0)$ is the integration variable. The energy term, $E(\xi)$ = $\sqrt{\xi^{2} + \Delta^{2}(t)}$, represents the quasiparticle excitation energy relative to the Fermi level, $\Delta(t)$ being the temperature-dependent superconducting gap as a function of the reduced temperature t = $T/T_{\rm c}$. In the isotropic s-wave BCS approximation, the gap function follows:
\begin{equation}
\Delta(t) = \text{tanh}[1.82((1.018(1/t))-1)^{0.51}].
\end{equation}

The electronic specific heat is then obtained as $C_{el} = t\frac{dS}{dt}$. Fitting the experimental data to the isotropic s-wave BCS model, shown in \figref{Fig2}(c), yields a superconducting gap value $\Delta(0)/k_{B}T_{\rm c}$ = 1.50(6). This value is slightly lower than the weak-coupling BCS limit, suggesting weakly coupled superconductivity in ZrIrGe.
For further verification of the nature of superconductivity in ZrIrGe, microscopic measurement techniques such as muon spin rotation and relaxation or tunnel-diode oscillation (TDO) measurements are essential.

\begin{figure*}
\includegraphics[width=2.0\columnwidth]{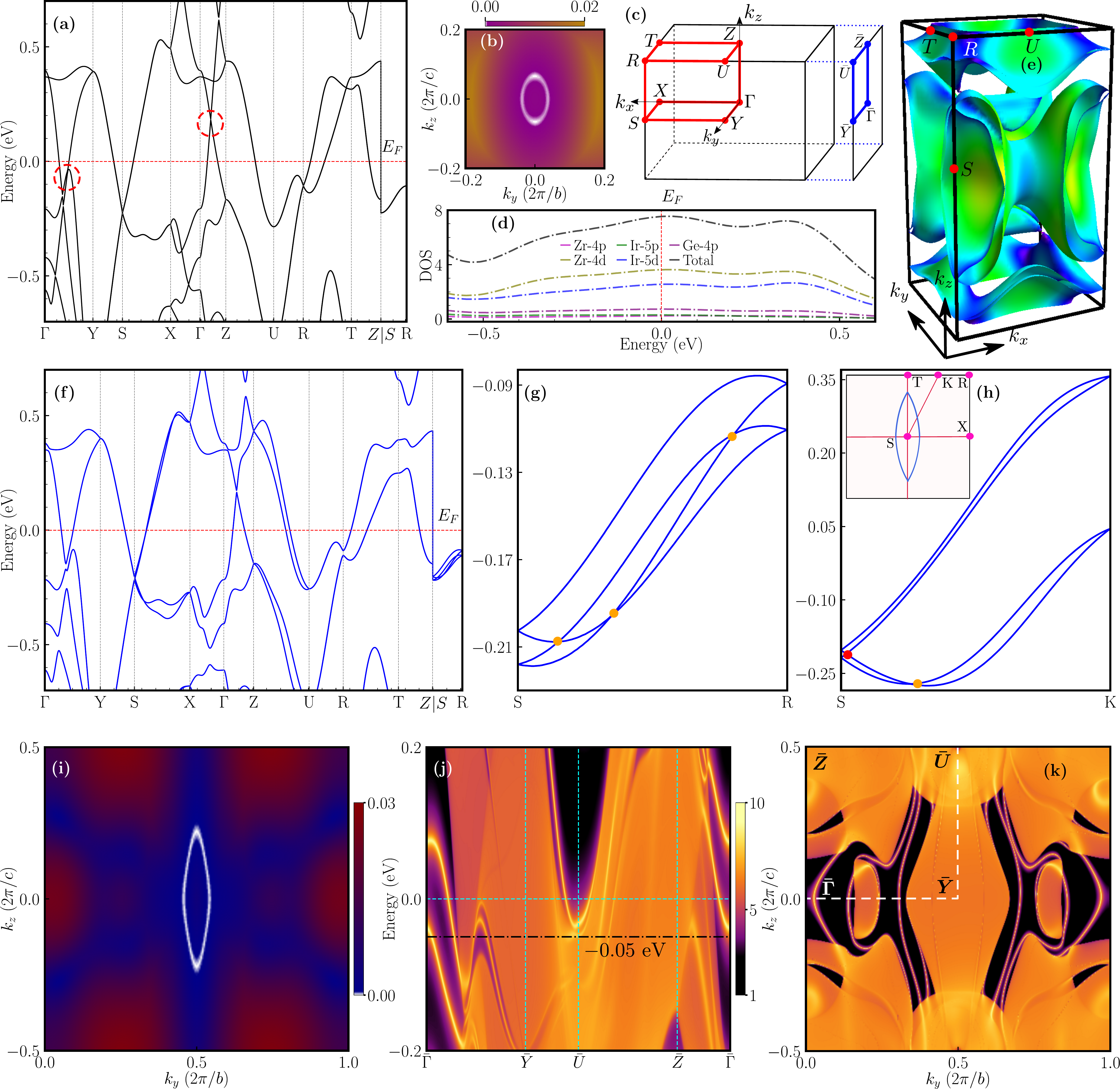}
\caption {\textbf{Electronic band structure, hourglass dispersions, Dirac rings and surface states of ZrIrGe:} (a) Bulk electronic band structure of ZrIrGe calculated in the absence of spin-orbit coupling (SOC). (b) Corresponding nodal ring centered at the $\Gamma$ point in the $k_x = 0$ plane. (c) The three-dimensional bulk Brillouin zone (BZ) and its projection onto the (100) surface, with high-symmetry points and paths marked by red dots and blue lines, respectively. (d) Orbital-resolved density of states of ZrIrGe without including spin-orbit coupling (SOC). (e) Fermi surface including SOC, showing multiple sheets across the Brillouin zone. (f) Electronic band structure with SOC included. (g,h) Hourglass-like band dispersions along the high-symmetry directions $S-R$, and $S-K$, where $K$ denotes the midpoint between $T$ and $R$. The inset in (h) schematically depicts a fourfold-degenerate Dirac ring, formed by neck points (red dots) of the hybrid hourglass dispersion on the $k_x = \pi$ plane. (i) Distribution of the hourglass Dirac ring (shown in white) encircling the $S$ point, with the color scale representing the local band gap. (j) Surface state spectrum along high-symmetry paths of the projected (100) two-dimensional surface Brillouin zone. (k) Constant-energy slice of the surface spectrum at $-0.05$ eV, revealing topological surface Fermi arcs.}
\label{fig:band_structure_topology}
\end{figure*}

\subsection{Electronic band structure from first principles calculations}
The electronic band structure of ZrIrGe exhibits distinctive topological features that originate from its nonsymmorphic crystal symmetry, governed by the orthorhombic space group Pnma (No. 62). This space group includes three essential symmetry operations: inversion symmetry $\mathcal{P}:(x, y, z) \rightarrow (-x, -y, z)$, mirror symmetry $\mathcal{M}_y:(x, y, z) \rightarrow (x, -y + \frac{1}{2}, z)$, and glide mirror symmetry $\mathsf{G}_x:(x, y, z) \rightarrow (-x + \frac{1}{2}, y + \frac{1}{2}, z + \frac{1}{2})$. These nonsymmorphic symmetries play a crucial role in enforcing protected band crossings and shaping the topological nature of the electronic states. To explore these features in detail, we performed first-principles electronic structure calculations based on density functional theory (DFT) within the generalized gradient approximation (GGA) using the Perdew-Burke-Ernzerhof (PBE) functional~\cite{Perdew1996}. The band structure without spin-orbit coupling (SOC), shown in \figref{fig:band_structure_topology}(a), reveals several dispersive bands crossing the Fermi level, resulting in multiple electron and hole pockets and underscoring the multiband character of ZrIrGe. The low-energy electronic states near the Fermi level are predominantly contributed by the Zr-4d, Ir-5d, Ge-4p, Ir-5p, and Zr-4p orbitals, in descending order-consistent with the orbital resolved projected density of states shown in \figref{fig:band_structure_topology}(d). Importantly, multiple band crossing points appear along the $\Gamma-Y$ and $\Gamma-Z$ directions, as highlighted by red circles. These crossings are protected by the glide mirror symmetry $\mathsf{G}_x$ and collectively form a nodal ring encircling the $\Gamma$ point in the $k_x = 0$ plane, as illustrated in \figref{fig:band_structure_topology}(b). When SOC is included, this nodal loop gaps out, leading to band inversion near the $\Gamma$ point and giving rise to nontrivial topology in ZrIrGe. Although the inclusion of SOC does not open a global band gap across the entire Brillouin zone, a well-defined $\mathbb{Z}_2$ topological invariant can still be defined in the $k_y-k_z$ plane. Our calculations reveal a nontrivial topological index of $\mathbb{Z}_2=1$, confirming the presence of robust surface states on the (100) surface. These surface states are clearly separated from the bulk bands and disperse across the Fermi level, underscoring the nontrivial topological nature of ZrIrGe (see \figref{fig:band_structure_topology}(j)).

The band structure with SOC, shown in \figref{fig:band_structure_topology}(f), exhibits pronounced band splitting near the Fermi level, particularly along the $R-T$ direction, with a maximum splitting of approximately {\color{red} $140$} meV. This effect arises from the strong spin-orbit interaction associated with the dominant Zr-4d and Ir-5d orbital contributions. The combined Fermi surface of ZrIrGe, shown in \figref{fig:band_structure_topology}(e), reveals nearly parallel sheets extending across the Brillouin zone-an electronic structure conducive to strong interband superconducting pairing~\cite{ghosh2022time,Weng2016}.

Additionally, the SOC-included band structure of ZrIrGe exhibits distinctive hourglass-shaped dispersions~\cite{gao2020r,Li2018} along the high-symmetry directions $S-R$, $S-X$, and $S-K$ (where $K$ denotes the midpoint between $T$ and $R$), as illustrated in \figref{fig:band_structure_topology}(f). An hourglass dispersion is a symmetry-enforced band connectivity in energy-momentum (E-k) space, typically arising in crystals with nonsymmorphic glide mirror symmetry $\mathsf{G}_x$ in combination with time-reversal symmetry. In such systems, the eigenstates at different high-symmetry momenta belong to distinct irreducible representations, which necessitates partner switching of Kramers-degenerate bands along certain high-symmetry lines. This constraint forces the bands to cross at a symmetry-protected degeneracy point-the ``neck'' of the hourglass-before rehybridizing, producing a dispersion profile that resembles the shape of an hourglass. Close-up views of these hourglass dispersions near the Fermi level-protected by the glide mirror symmetry $\mathsf{G}_x$, are presented in \figref{fig:band_structure_topology}(g,h). In addition to these hourglass features, further analysis reveals the presence of a Dirac ring centered around the $S$ point within the $k_x = \pi$ plane (as shown in the inset of \figref{fig:band_structure_topology}(h)), accompanied by topologically nontrivial surface states.

To investigate the origin of hourglass dispersions~\cite{gao2020r,Li2018}, we begin by examining the hourglass dispersion along the $S-R$ high symmetry path, which corresponds to the line ($\pi,\pi,k_z$) with $-\pi < k_z < \pi$. This path lies within the $k_x=\pi$ plane and remains invariant under the glide mirror operation $\mathsf{G}_x$. Along this path, the square of the glide operation satisfies $\mathsf{G}^2_x=\mathcal{T}_{011}\Bar{E}=e^{-ik_z}$, where $\Bar{E}$ denotes a $2\pi$ spin rotation and $\mathcal{T}_{011}$ is a lattice translation mapping $(x,y,z)$ to $(x,y+b,z+c)$. Consequently, the eigenvalues $g_x$ of $\mathsf{G}_x$ must take the form $\pm e^{-ik_z/2}$. For a Bloch state $\ket{\phi_n}$, its Kramers partner $\mathcal{P}\mathcal{T}\ket{\phi_n}$ shares the same glide eigenvalue: $\mathsf{G}_x(\mathcal{PT}\ket{\phi_n})=g_x(\mathcal{PT}\ket{\phi_n})$. At the time-reversal-invariant momenta (TRIM) points $S$ and $R$, this condition also holds for the states $\mathcal{T}\ket{\phi_n}$ and $\mathcal{P}\ket{\phi_n}$, due to the centrosymmetric nature of the crystal structure ZrIrGe.

At the $S$ point $(\pi, \pi, 0)$, the four states $\ket{\phi_n}$, $\mathcal{T}\ket{\phi_n}$, $\mathcal{P}\ket{\phi_n}$, and $\mathcal{PT}\ket{\phi_n}$ form a degenerate quartet with glide eigenvalues $g_x=\pm 1$, where the states with $g_x=-1$ typically reside at lower energy than those with $g_x=+1$. In contrast, at the $R$ point $(\pi, \pi, \pi)$, the glide eigenvalues become $g_x=\pm i$, and each Kramers pair consists of states with opposite glide eigenvalues, resulting in a quartet composed of two states with $g_x=+i$ and two with $g_x=-i$. This change in eigenvalue structure between the $S$ and $R$ points enforces a symmetry-protected band crossing along the $S-R$ path, giving rise to a characteristic hourglass dispersion. Since this crossing is mandated by nonsymmorphic glide mirror symmetry and does not rely on a band inversion mechanism, it is highly robust. A similar symmetry-based argument applies to the $S-X$ and $S-K$ high-symmetry paths, which also host glide mirror symmetry-protected hourglass dispersions. Notably, the hourglass features along $S-X$ and $S-K$ extend across a wide energy range and cross the Fermi level, while those along $S-R$ are confined to a narrower energy window just below the Fermi level.

The hourglass dispersions of ZrIrGe, shown in \figref{fig:band_structure_topology}(g,h), also feature both type-I and type-II Dirac points, marked by red and orange dots, respectively. Type-I Dirac points, also known as neck points, occur along any $S-K$ path, where $K$ is an arbitrary point along the high-symmetry $T-R$ direction. A similar behavior is observed along the $R-X$ path. These symmetry-protected neck points (red dots) collectively form a continuous Dirac nodal ring centered around the $S$ point in the $k_x = \pi$ plane, as schematically illustrated in the inset of \figref{fig:band_structure_topology}(h). This distinctive hourglass-type Dirac ring arises directly from the nonsymmorphic space-group symmetry and is confirmed by first-principles DFT calculations. The corresponding DFT results for ZrIrGe are shown in \figref{fig:band_structure_topology}(i).

Finally, we examine the surface electronic structure of ZrIrGe by projecting the bulk bands onto the (100) surface Brillouin zone, as shown in \figref{fig:band_structure_topology}(j). The surface spectrum reveals several topologically nontrivial surface states crossing the Fermi level. These states originate from bulk Dirac nodal loops, which project onto finite areas of the surface Brillouin zone, producing characteristic drumhead-like surface features~\cite{Li2018}. Owing to the broken inversion symmetry at the surface, these drumhead states exhibit a splitting~\cite{Li2018}. These surface bands, intimately connected to the bulk Dirac rings, appear near energy $\sim -0.05$ eV, as illustrated in \figref{fig:band_structure_topology}(k).

\subsection{Electronic properties and Uemura plot}
The electronic properties of ZrIrGe, including the electronic mean free path ($l_{e}$), effective mass ($m^{*}$), and London penetration depth ($\lambda_{L}$), were determined by simultaneously solving a set of equations that assume a spherical Fermi surface. The relevant equations are \cite{mayoh2017superconducting}:
\begin{equation}
\gamma_{n} = \left(\frac{\pi}{3}\right)^{2/3}\frac{k_{B}^{2}m^{*}V_{\mathrm{f.u.}}n^{1/3}}{\hbar^{2}N_{A}};
\label{eqn17:gf}
\end{equation}
\begin{equation}
\textit{l}_{e} = \frac{3\pi^{2}{\hbar}^{3}}{e^{2}\rho_{0}m^{*2}v_{\mathrm{F}}^{2}}, n = \frac{1}{3\pi^{2}}\left(\frac{m^{*}v_{\mathrm{F}}}{\hbar}\right)^{3};
\label{eqn18:le,n}
\end{equation}
where $k_{B}$ is the Boltzmann constant, $V_{f.u.}$ is the volume of the formula unit, and $N_{A}$ is Avogadro's number. 

For superconductors in the dirty limit, the effective Ginzburg-Landau penetration depth ($\lambda_{GL}$(0)) is related to the London penetration depth ($\lambda_{L}$(0)) by: 
\begin{equation}
\lambda_{GL}(0) = 
\lambda_{L}
\left(1+\frac{\xi_{0}}{\textit{l}_{e}}\right)^{1/2}, \lambda_{L} =
\left(\frac{m^{*}}{\mu_{0}n e^{2}}\right)^{1/2}
\label{eqn20:f}
\end{equation}
\begin{equation}
\frac{\xi_{GL}(0)}{\xi_{0}} = \frac{\pi}{2\sqrt{3}}\left(1+\frac{\xi_{0}}{\textit{l}_{e}}\right)^{-1/2}
\label{eqn21:xil}
\end{equation}
where $\xi_{0}$ is the superconducting coherence length. A high value of $\xi_{0}/l_{e}$ confirms that ZrIrGe is in the dirty limit. The calculated electronic parameters are summarized in \tableref{tbl: parameters}. 

\begin{figure}
\includegraphics[width=0.995\columnwidth]{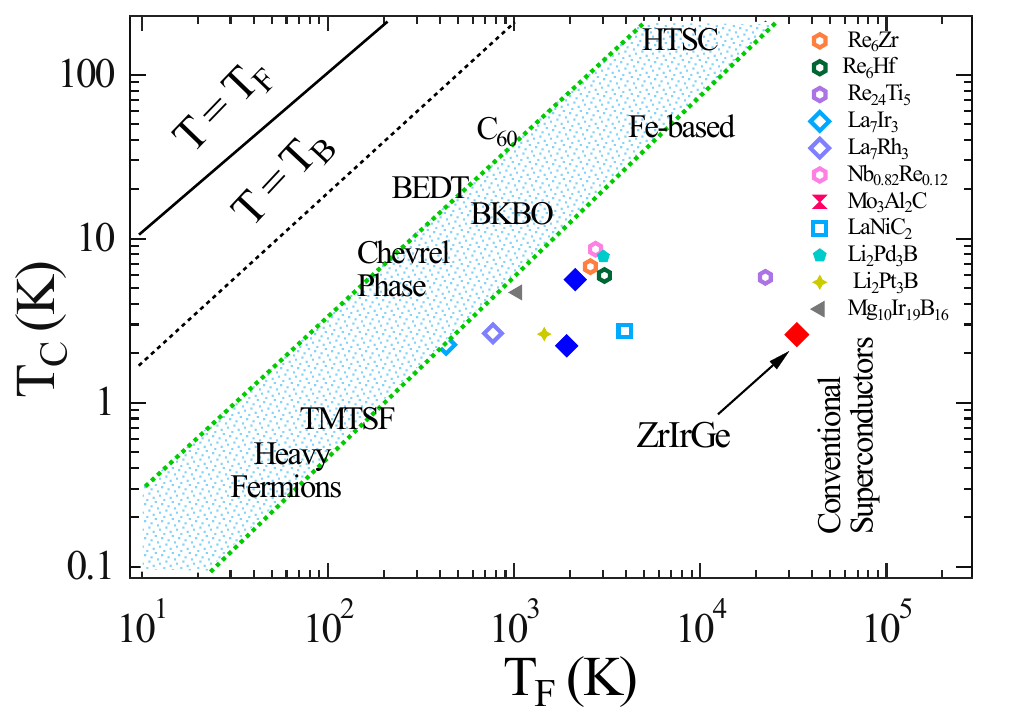}
\caption {\label{Fig4} The Uemura plot between T$_{c}$ and T$_{F}$ for ZrIrGe is indicated by a red diamond marker positioned near the conventional superconductivity range.}
\end{figure}

To further analyze the nature of superconductivity in ZrIrGe, we employed the Uemura classification scheme. This scheme categorizes superconductors based on the ratio of their ($T_{\rm c}$) to the Fermi temperature (T$_{F}$). Superconductors with 0.01 $\le$ $\frac{T_{\rm c}}{T_{F}}$ $\le$ 0.1 are classified as unconventional superconductors. The Fermi temperature (T$_{F}$) can be estimated using single-band model \cite{hillier1997classification};
\begin{equation}
k_{B}T_{F} = \frac{\hbar^{2}}{2}(3\pi^{2})^{2/3}\frac{n^{2/3}}{m^{*}}. 
\label{Eq13:tf}
\end{equation}
This expression provides only an approximate, order-of-magnitude estimate rather than a precise determination, particularly given the multiband nature of ZrIrGe. By substituting our calculated values for $n$ and $m^{*}$, we obtain $T_{F}$ = 33026 K. The corresponding T$_{c}$/T$_{F}$ ratio is 0.000078, which places ZrIrGe outside the unconventional superconducting regime [see the red diamond marker in Fig. \ref{Fig4}].

\begin{table}[htbp!]
\caption{Normal and superconducting parameters of ZrIrGe derived from magnetisation, resistivity and specific heat measurements.}
\label{tbl: parameters}
\setlength{\tabcolsep}{15pt}
\begin{center}
\begin{tabular}{p{0.17\linewidth}p{0.25\linewidth}p{0.17\linewidth}}
\hline
\hline
Parameter &Units & ZrIrGe \\
\hline
T$_{c}$ & K & 2.84(7) \\
H$_{c1}(0)$ & mT & 11.2(1) \\ 
H$_{c2}^{mag}$(0) & T & 0.38(1) \\
H$_{c2}^{res}$(0) & T & 0.42(1) \\
H$_{c2}^{P}$(0) & T & 5.29(5) \\
H$_{c}$ & mT & 49.8(1) \\
$\xi_{GL}$& $nm$ & 29.4(4) \\
$\lambda_{GL}^{mag}$& $nm$ & 163.9(3) \\
$k_{GL}$& & 5.5(5) \\
$RRR$ & &7.31(5)\\
$\gamma_{n}$& mJ/(mol K$^{2}$) & 8.25(8) \\
$\theta_{D}$& K& 374(9)\\
$\frac{\Delta(0)}{k_{B}T_{\rm c}}$ (sp) &   & 1.50(6) \\
$\lambda_{e-ph}$ &  &0.51(3) \\
D$_{C}$(E$_{f}$) & states/(eV f.u.) &3.50(1)\\
n & $10^{28} m^{-3}$ &1.4(9)\\
$m^{*}$ & -&0.77(7) \\
$v_{F}$ & $10^{5}$ $ms^{-1}$ &11.3(6)\\
$\xi_{0}/l_{e}$ & - & 17.3(3)\\
T$_{F}$ & $10^{4}$ K &3.30(2)\\
T$_{c}/$T$_{F}$ & &0.000078\\
\hline
\hline
\end{tabular}
\par\medskip\footnotesize
\end{center}
\end{table}


\section{Summary and Conclusion}

In summary, we have comprehensively investigated the superconducting and topological properties of ZrIrGe through a combination of experimental measurements and first-principles calculations. Synthesized by arc melting, ZrIrGe adopts an orthorhombic TiNiSi-type structure (space group $Pnma$) characterized by three glide-plane symmetries. The presence of heavy $5d$ Ir atoms gives rise to strong spin-orbit coupling (SOC), which plays a pivotal role in shaping the electronic structure. Our first-principles analysis reveals hourglass-shaped band dispersions along several high-symmetry directions, protected by glide mirror symmetry. The neck points of these hourglass-shaped dispersions form a fourfold-degenerate Dirac nodal ring within the $k_x = \pi$ plane, centered at the $S$ point. On the (100) surface, we further identify topologically nontrivial surface states, including Fermi arcs that link the projections of the bulk Dirac ring, as well as surface states crossing the Fermi level near the $\Gamma$ point, which originate from a band inversion. These features make ZrIrGe a promising platform for studying topological superconductivity, and they are readily amenable to experimental probes such as ARPES and STM/STS~\cite{Chen2022}.

Transport and thermodynamic measurements provide clear evidence for bulk type-II superconductivity in ZrIrGe, with a superconducting transition at $T_c = 2.84(7)$ K. Normal-state resistivity confirms metallic behavior, while magnetization and specific heat data indicate a fully gapped, weakly coupled superconducting state consistent with the weak-coupling BCS framework. The extracted superconducting parameters, including the electron-phonon coupling constant ($\lambda_{e\text{-}ph}$), lie between those of TiIrGe and HfIrGe \cite{meena2025topological}, following a similar trend to that observed in the structurally analogous MIrSi (M = Ti, Zr, Hf) compounds \cite{kase2016superconductivity}. These results collectively establish ZrIrGe as a compelling candidate for realizing intrinsic topological superconductivity, offering a stoichiometric and clean material system where symmetry-enforced topological surface states coexist with conventional superconductivity.

\section{Acknowledgments}
P.~K.~M. acknowledges the funding agency Council of Scientific and Industrial Research, Government of India, for providing the SRF fellowship award no. 09/1020(0174)/2019-EMR-I. R.~P.~S. acknowledges the SERB, Government of India, for the Core Research Grant no. CRG/2023/000817. SKG also acknowledges financial support from SERB, Government of India via the Startup Research Grant: SRG/2023/000934 and IIT Kanpur via the Initiation Grant (IITK/PHY/2022116). DS and SKG utilized the \textit{Andromeda} server at IIT Kanpur for numerical calculations.

\bibliography{ref}

\end{document}